\documentclass[11pt,letterpaper]{article}
\usepackage[T1]{fontenc}
\usepackage[utf8]{inputenc}
\usepackage{lmodern}
\usepackage[margin=1in]{geometry}
\usepackage{amsmath,amssymb}
\usepackage{graphicx}
\usepackage{longtable,booktabs,array,calc}
\usepackage{etoolbox}
\usepackage{microtype}
\usepackage[section]{placeins}
\usepackage[font=small,labelfont=bf]{caption}
\usepackage{upquote}
\usepackage{fancyvrb}
\usepackage{fvextra}
\DefineVerbatimEnvironment{Highlighting}{Verbatim}{breaklines,commandchars=\\\{\}}

\usepackage[numbers,sort&compress]{natbib}
\usepackage{xurl}
\usepackage[hidelinks]{hyperref}
\hypersetup{pdftitle={When Agent Metrics Measure Different Things: An
Evidence-Grounded Audit of the Praxa AI
Pipeline},pdfauthor={Stefan G. Creadore; Peyton Woakz},pdfsubject={Retrospective agent-system measurement audit}}

\AtBeginEnvironment{longtable}{\footnotesize}

\makeatletter
\def\maxwidth{\ifdim\Gin@nat@width>\linewidth\linewidth\else\Gin@nat@width\fi}
\def\maxheight{\ifdim\Gin@nat@height>0.72\textheight0.72\textheight\else\Gin@nat@height\fi}
\makeatother
\setkeys{Gin}{width=\maxwidth,height=\maxheight,keepaspectratio}
\title{\vspace{-1.0cm}When Agent Metrics Measure Different Things: An
Evidence-Grounded Audit of the Praxa AI Pipeline}
\author{Stefan G. Creadore\thanks{Corresponding author: \href{mailto:Stefan@praxa.io}{Stefan@praxa.io}. ORCID: \href{https://orcid.org/0000-0003-2268-053X}{0000-0003-2268-053X}.}\\[2pt]
\small Tampa, FL, USA
\and Peyton Woakz}
\date{10 September 2026}
\begin{document}
\maketitle
\begin{center}\small
\href{https://praxa.io}{praxa.io}\quad\textbar\quad
\href{https://github.com/praxa-labs/}{github.com/praxa-labs/}
\end{center}
\begin{abstract}
Agent evaluations can be numerically correct while measuring a different
construct from the one implied by their labels. We present a
retrospective measurement audit of selected Praxa AI implementation
files, historical evaluation artifacts, and operational records. A
139-case offline routing report contains 112 passes and 27 failures
despite zero gating failures, because known gaps are explicitly exempted
from the gate. An identifier-free export represents 8,843 tool-attempt
rows: 8,395 recorded durations and 448 missing values. Of the durations,
121 equal the signed 32-bit maximum and carry abandoned-client labels;
inspected database code clamps elapsed lifecycle age. The pooled
recorded 99th percentile is 2,147,483,647 ms, versus 38,118.31 ms among
server-observed completed calls. This is a stratum contrast, not a
treatment effect. In a documented single-trajectory compaction pilot,
the reported follow-up input reduction is 94.39\%, but the reduction
across the trigger and follow-up calls together is 46.54\%. We reproduce
the descriptive calculations, verify 91 timing statistics through a
separate weighted rational-arithmetic implementation, and execute 13
scoring-function and 12 analysis-verifier tests. Finite-completion
bounds show how missing durations limit all-row timing statements
without imputing values. The contribution is a source-linked case study
and reusable verification package for separating gate policy, lifecycle
timing, and request-level accounting from broader agent-performance
claims. Historical provider runs and the full current pipeline were not
independently reproduced; general capability superiority and
population-level statistical significance are not established.
\end{abstract}
\section{Introduction}\label{introduction}

An agent may select an appropriate tool, obtain a terminal tool status,
and still fail to accomplish the user's task. Conversely, a request may
wait for a person or remain unresolved long after active execution has
stopped. A useful evaluation must specify which of these events it
measures. More observations do not resolve a mismatch between the
recorded variable and the scientific question.

Prior work has identified weaknesses in agent evaluation that include
incomplete cost accounting, inadequate holdouts, and confusion between
the needs of model and application developers \citep{kapoor2024}.
Task-oriented benchmarks such as \(\tau\)-bench evaluate a different
object from a software gate: they compare resulting database state with
an annotated task goal and examine reliability across repeated trials
\citep{yao2024}. These distinctions motivate a narrow question for the
present study: what do the available Praxa measurements actually
establish?

Praxa is the system named in the inspected repository \citep{praxarepo}.
Rather than assume that its architecture supplies a new learning
capability, we trace selected quantitative claims to their measurement
implementations, source artifacts, and denominators. The audit yields
three concrete cases. A passing development gate coexists with failing
acceptance cases. A latency column includes elapsed ages of abandoned
client calls. A large follow-up token reduction becomes smaller when its
trigger call is included. Each case is informative precisely because its
arithmetic can be correct while a broader interpretation is unsupported.

The contribution is a reproducible measurement case study with three
parts. \textbf{C1: measurement-boundary evidence.} We connect gate
exemptions, a clipped lifecycle-age field, and selective token
accounting to explicit source artifacts and finite denominators.
\textbf{C2: computational verification.} We provide a hashed timing
multiset, executable tests of the original scoring function, and a
separate weighted-arithmetic verifier; derived comparison totals and
plots share the same source inputs. \textbf{C3: bounded interpretation
of missing records.} We show how the observed missingness limits finite
all-row threshold statements without imputing durations. These are
contributions of evidence and reproducible application, not claims to
have invented gating distinctions, partial-identification bounds, or
measurement science.

We also propose a field-level measurement contract and controlled
follow-up studies. Their operational effectiveness has not been
evaluated. No new learning algorithm, state-of-the-art task performance,
or experimentally established advantage over retrieval-augmented
generation is claimed.

\section{Scope, research questions, and evidence
standard}\label{scope-research-questions-and-evidence-standard}

\subsection{What was inspected}\label{what-was-inspected}

The repository review was pinned to main commit
\nolinkurl{83d29cf51137d57ad11e5ad54ec2aea6f243b90d}, timestamped 9
September 2026 at 22:27:05 UTC (E01). We inspected selected
routing-evaluation code, scoring contracts, benchmark documentation, and
architecture/observability documentation. The historical routing report
identifies a different base revision,
\nolinkurl{e66dc504f5207d6174626f89270e6d1677927ef3} (E02). A base
revision recorded in a report does not by itself prove that the
generating working tree was clean. Database procedure metadata was
inspected independently; its state was not proven to match either Git
revision.

The original read-only database queries returned operational aggregates
and evaluation-table counts. The additional manuscript review reuses
those dated extracts; it does not silently refresh the operational
population. Direct access to Langfuse traces and Cloudflare operational
APIs was not available in this session after connector discovery
attempts. Their repository integration documents were reviewed instead.
Repository search and selected table inspection do not establish an
exhaustive inventory of the organization's experiments. In particular,
an unavailable trace or an empty selected table is not evidence that a
study never occurred.

We did not run the complete application, the historical 139-case routing
harness, a live provider benchmark, or a deployed task-completion test.
No production data was modified; no sweep, finalization, deployment, or
external purchasing/messaging function was invoked. The original audit
executed read-only extraction, descriptive reanalysis, and isolated
software conformance tests. This revision adds exact payload-to-table
reconciliation, a separately implemented arithmetic check, additional
software tests, and missingness sensitivity calculations. These are new
checks of retained evidence, not new agent trajectories.

\subsection{Research questions}\label{research-questions}

\textbf{RQ1: What separates case correctness from development gating?}
We examine whether the historical case total, category totals, known-gap
exemptions, and implemented denominators agree. This is a
finite-artifact consistency question, not a test of a population success
probability.

\textbf{RQ2: What event does the recorded latency measure?} We inspect
missingness, saturation, reporting authority, terminal status, and
relevant database logic. We test the descriptive proposition that the
retained frequency data reproduce the queried counts and quantiles. We
do not treat authority/status groups as randomized arms. We also ask
which finite all-row threshold statements remain possible when some
recorded durations are missing.

\textbf{RQ3: How sensitive is the documented compaction comparison to
its accounting boundary?} We recompute the difference between
follow-up-only input consumption and the sum of the two reported calls.
The numerical inputs are historical documentary evidence, not
independently rerun provider measurements.

\textbf{RQ4: Which stronger claims remain unresolved?} We identify the
additional controls needed to evaluate current-path tool admission,
verified completion, durable effects, and compaction efficiency. These
questions define prospective work rather than retroactively creating
hypotheses for experiments that were not run.

For RQ1--RQ3, null and alternative hypotheses about a sampled population
are not defined: the analysis concerns the observed finite artifacts.
Statistical testing would require a sampling model that the retained
evidence does not supply. Explicit hypotheses and controls for future
comparisons appear in Section 8.

\subsection{Evidence classes and claim
discipline}\label{evidence-classes-and-claim-discipline}

Evidence identifiers E01--E15 link the manuscript to the supplementary
registry. We distinguish observed source facts, newly executed
measurements, historical source-reported results, and proposed
experiments. A claim can be verified as a calculation without its source
experiment being independently reproduced. For example, a token
reduction calculated from two documented counts has reproducible
arithmetic but remains conditional on the accuracy and meaning of those
counts.

The accompanying claims ledger uses VERIFIED, SUPPORTED, PRELIMINARY,
HYPOTHESIS, NOT ESTABLISHED, and REJECTED as admissible classifications.
Primary conclusions here are confined to verified descriptive
calculations and explicitly qualified source-supported interpretations.
``Not established'' describes the limits of this audit; it is not a
finding that a capability is impossible or absent.

\section{System context and related
work}\label{system-context-and-related-work}

\subsection{Relevant implementation
boundaries}\label{relevant-implementation-boundaries}

Architecture documentation describes an Expo application, routed
Cloudflare Workers, Supabase persistence, and additional
state/background-execution surfaces (E11). This is documentary context,
not a deployed-topology certification. The empirical analysis focuses on
the smaller boundaries in Table 1, for which the retrieved evidence can
be specified precisely.

\begin{table}[!htbp]
\centering
\footnotesize
\caption{Selected system boundaries and their evidential scope. A
documented or code-defined interface is not an experimentally validated
capability.}
\vspace{4pt}
\begin{tabular}{@{}>{\raggedright\arraybackslash}p{(\columnwidth - 4\tabcolsep) * \real{0.25}}
>{\raggedright\arraybackslash}p{(\columnwidth - 4\tabcolsep) * \real{0.4}}
>{\raggedright\arraybackslash}p{(\columnwidth - 4\tabcolsep) * \real{0.35}}@{}}

\toprule\noalign{}
\begin{minipage}[b]{\linewidth}\raggedright
Boundary
\end{minipage} & \begin{minipage}[b]{\linewidth}\raggedright
Retrieved evidence
\end{minipage} & \begin{minipage}[b]{\linewidth}\raggedright
What it can establish
\end{minipage} \\
\midrule\noalign{}

Routing and admission & Offline report, metric implementation, runner
excerpt (E02--E03) & Asserted pre-model routing/scoping behavior and
metric definitions \\
Tool-attempt persistence & Frequency aggregates, status/authority
diagnostics, SQL excerpts (E06--E08) & Recorded attempt values and
selected lifecycle semantics \\
Provider search/compaction & Historical pilot document (E04) & Reported
local comparison, conditional arithmetic, and stated caveats \\
Effect verification & Frontier scorer and contracts (E10) & Declared
scoring requirements; not deployed compliance \\
Observability & Langfuse integration document (E12) & Intended trace
fields and prompt-block boundaries; not live trace results \\
\bottomrule
\end{tabular}
\end{table}

The offline routing harness is particularly easy to overinterpret. Its
documentation specifies no model or network calls, name-oriented
admission tests, and a pre-V2 evaluation path (E02--E03). The flag
document at the audited revision describes ordinary chat as V2 and calls
its former selector flag obsolete (E11). Thus the historical report
cannot be promoted to a benchmark of the provider-visible tool set at
that revision. Follow-up cases also do not establish full
conversational-history processing merely because their category name
contains ``follow-up.''

The layered Frontier scorer distinguishes evidence classes and mandatory
layers. Its original contract has 16 mandatory layers, while the
candidate version adds an explicit task-graph layer. The inspected
baseline scorer sets \nolinkurl{candidateCompletionEligible} to false
(E10). This is a useful separation between a baseline rubric and
candidate-level proof; counting named layers would not show that a live
task satisfied them.

\subsection{Position relative to prior
work}\label{position-relative-to-prior-work}

Our analysis concerns measurement validity rather than downstream
benchmark rankings. Kapoor et al.~motivate joint attention to cost,
accuracy, and evaluation design \citep{kapoor2024}. \(\tau\)-bench's
state-based outcome checking illustrates why a terminal tool label and a
verified user goal should be separate variables \citep{yao2024}.
AgentRewardBench studies the reliability of automatic trajectory
evaluation, which is relevant when score labels are mistaken for
independent outcome evidence \citep{lu2025}. We do not transfer any of
these papers' quantitative results to Praxa. Unlike those benchmark
studies, our unit of analysis is a retained measurement artifact or a
recorded field, rather than a newly executed agent task.

Recent harness research supplies useful methodological comparisons.
HarnessDev separates harness creation from evolution and tests held-out
capability, efficiency, and dependence on the executing model
\citep{wu2026}. Repo-To-Skill evaluates reusable operating knowledge
with the backbone, harness, and downstream budget held fixed
\citep{chen2026}. AutoResearch includes evidence-based review before
accepting experimental conclusions \citep{ren2026}. These studies show
that controlled harness evaluation and evidence review already have
substantial prior art. We claim neither to originate those ideas nor to
have replicated their evaluations.

Other recent reference papers address distinct constructs. AI Can
Learn Scientific Taste studies learned scientific judgment and ideation,
including generalization beyond training conditions \citep{tong2026}.
Heterogeneous Agent Collaborative Reinforcement Learning concerns shared
verified rollouts during training with independent inference
\citep{zhang2026}. Omni Interaction Agent reports a streaming multimodal
architecture and evaluations across interaction and agentic dimensions
\citep{orantqing2026}. They are contextual references, not supporting
evidence for Praxa outcomes and not matched baselines for the artifacts
audited here. No scientific-taste, collaborative-training, or
multimodal-interaction capability is measured in the present study.

\section{Methods and reproducibility
design}\label{methods-and-reproducibility-design}

\subsection{Historical aggregate
extraction}\label{historical-aggregate-extraction}

For the acceptance report, we retained its complete retrieved summary,
nine category summaries, explicit rate denominators, generation time,
historical base SHA, and source blob identity. We did not reconstruct
missing individual cases from those aggregates. Integer rate numerators
in the CSV are derived from each reported rate and its explicit
denominator and checked against the overall totals. Rounded category
means are retained as reported, not expanded into fictional per-case
observations.

For the provider pilot, we retained the document's per-arm values,
reported software versions, sample sizes, and stated reinjection/cost
limitations. Discovery percentiles are source-reported summaries of two
executions per arm; the raw execution timings were not recovered. We
therefore do not infer individual run values, a percentile convention,
standard deviations, or confidence intervals from them.

The ledger-outage comparison is likewise documentary. We retain its
named baseline, uncommitted candidate designation, seven-fixture
denominator, and reported Wilson intervals. Only the interval arithmetic
is reproduced. The absent immutable candidate code and raw fixture
provenance prevent an environment-identical replay. The absence of a
verified sampling or randomization design, rather than the arithmetic
itself, limits inferential interpretation.

\subsection{Identifier-free operational
extraction}\label{identifier-free-operational-extraction}

The timing extract groups \nolinkurl{agent_tool_call_attempts} by
reporting authority, terminal status, and exact integer
\nolinkurl{latency_ms}, retaining each value's frequency. Null values
occupy explicit missing-value bins. It includes all rows returned for
this table at extraction, without a success-only filter or removal of
slow observations. Its row-creation timestamps span 21 July through 9
September 2026; this does not prove complete historical coverage or
distinguish production traffic from every possible test/backfill source
(E06--E07).

Frequency payloads were retrieved in separate read-only queries between
23:02:45.811561 and 23:11:54.839382 UTC on 9 September. Each compact
payload carries a PostgreSQL-returned MD5 digest; the local copy
reproduces that digest. In the revision, every parsed non-null
value-frequency cell is also checked against the analysis CSV, not
merely against a total. These checks establish identity between the
retained strings and the table used for plotting, not independent
attestation of the originating database. The extracts were not a single
repeatable-read transaction. Their totals and quantiles reconcile with
separately queried summaries, but that agreement does not prove atomic
snapshot consistency.

No conversation text, owner identifier, tool-call identifier, or raw
account content is included in the release candidate. Frequency
aggregation preserves the multiset needed for univariate descriptive
statistics while removing event order and unit linkage. It does not
preserve user/task clusters, repeated-experiment pairings, or the
evidence needed to estimate task completion. Identifier removal alone
is not a privacy guarantee.

\subsection{Statistics and
denominators}\label{statistics-and-denominators}

Let \(f_g(v)\) be the retained count of numeric duration \(v\) in
stratum \(g\). The number of observed durations and their mean are

\[
n_g=\sum_v f_g(v),\qquad
\bar{x}_g=\frac{\sum_v v f_g(v)}{n_g}.
\]

The sample variance reported in the supplement uses

\[
s_g^2=\frac{\sum_v f_g(v)(v-\bar{x}_g)^2}{n_g-1}.
\]

The term \emph{sample variance} identifies the denominator convention;
it does not assert independent sampling. Quantiles use linear
interpolation between ordered observations, matching the interpolation
definition of PostgreSQL \nolinkurl{percentile_cont} \citep{postgres17}.
For quantile level \(p\), let \(h=(n-1)p\), \(j=\lfloor h\rfloor\), and
\(a=h-j\). With zero-based ordered observations,

\[
Q(p)=(1-a)x_{(j)}+a x_{(\min(j+1,n-1))}.
\]

The supplementary table includes mean, median, sample standard
deviation, variance, minimum, maximum, quartiles, interquartile range,
and p50/p90/p95/p99. Nulls are counted but never imputed as zero.
Saturated values and observed zeros remain in the primary analysis. A
secondary deletion of values equal to the integer cap is explicitly a
sensitivity calculation, not a corrected execution-latency estimator.

Acceptance rates preserve their implemented denominators. Case pass rate
counts cases. Required-tool recall counts required tool-name assertions.
Forbidden-admission rate counts explicitly forbidden tool-name
assertions. Fast-path accuracy includes only cases with an explicit
fast-path expectation. Those denominators are not interchangeable, and
their counts are not independent additional agent trials.

A second implementation reads the frequency table without expanding it.
It uses integer sums and Python rational arithmetic for weighted means,
sample variances, and linear quantiles, then checks the primary
implementation's floating-point outputs to relative tolerance
\(10^{-12}\) and absolute tolerance \(10^{-6}\) in the reported units.
Square roots are checked as floating-point scalars. This diversifies the
calculation path; both implementations still depend on the same retained
source evidence.

To describe missingness without assuming random missingness, let \(n\)
be the number of observed durations, \(m\) the number missing, and
\(c(t)\) the number of observed durations no greater than threshold
\(t\). Across all assignments of below/above-threshold classifications
to the missing rows, the completed-table fraction obeys

\[
\frac{c(t)}{n+m} \leq F_{\mathrm{complete}}(t)
\leq \frac{c(t)+m}{n+m}.
\]

The endpoints are attained by putting all missing rows above or at/below
\(t\), respectively. This elementary bound is a finite sensitivity
calculation, not a confidence interval, an imputation, or proof that
every missing duration is recoverable or well-defined. Thresholds of 1,
5, 30, and 60 seconds are illustrative analysis choices, not
prespecified service objectives. Appendix D reports all selected
thresholds.

\subsection{Newly executed scoring
tests}\label{newly-executed-scoring-tests}

We copied the original \nolinkurl{computeAgenticMetrics} source and
verified its Git blob SHA,
\nolinkurl{01bf2308619a9c9b4a84f7ce5b6803d3942b6498}. An isolated Node
22.16.0 harness transpiled it with TypeScript 5.8.3 and executed 13
deliberately constructed software fixtures (E13). A dependency stub
throws if the stateless-tool registry is used; it was never called. The
test therefore exercises the original metric function without silently
replacing its scoring logic.

This setup is not the repository's full build. In particular, the
repository's stated TypeScript requirement differs from the transpiler
used for this isolated test, and neither project type-checking nor the
complete Bun routing runner was performed. The receipt records the exact
environment, source hash, fixtures, and outcome. A separate verifier
executes 12 deliberately constructed tests of duplicate detection, null
handling, quantile interpolation, cap retention, and bound endpoints
(E15). Deliberately altered test inputs are confined to verifier tests;
retained empirical files are not altered. No random seed applies to
these deterministic fixtures, and no model was invoked.

\section{Results}\label{results}

\subsection{Offline routing: a gate is not the case
denominator}\label{offline-routing-a-gate-is-not-the-case-denominator}

The historical report contains 139 cases, of which 112 pass and 27 fail:
\textbf{80.58\% case pass rate}. It also reports \textbf{zero gating
failures}. All 27 failures are known-gap failures exempted from the
development gate but retained in the metric denominator (E02--E03). The
two results are consistent with the documented policy. Reporting the
gate alone as a perfect evaluation score would change the meaning of the
result.

\begin{table}[!htbp]
\centering
\footnotesize
\caption{Historical acceptance categories (E02; reanalysis E14).
Percentages describe this curated corpus, not sampled user tasks. The
last row is the complement of the legacy category and is not an
independent benchmark.}
\vspace{4pt}
\begin{tabular}{@{}lrrrr@{}}

\toprule\noalign{}
Category & Pass & Total & Fail & Pass (\%) \\
\midrule\noalign{}

legacy & 63 & 63 & 0 & 100.00 \\
no tool & 12 & 17 & 5 & 70.59 \\
single tool & 13 & 16 & 3 & 81.25 \\
ambiguous family & 6 & 13 & 7 & 46.15 \\
tool search & 6 & 6 & 0 & 100.00 \\
multi tool & 5 & 6 & 1 & 83.33 \\
tool name mention & 4 & 7 & 3 & 57.14 \\
follow up & 1 & 5 & 4 & 20.00 \\
multi intent & 2 & 6 & 4 & 33.33 \\
All cases & 112 & 139 & 27 & 80.58 \\
Nonlegacy only & 49 & 76 & 27 & 64.47 \\
\bottomrule
\end{tabular}
\end{table}

\begin{figure}[!htbp]
\centering
\includegraphics[width=0.95\textwidth,height=\textheight]{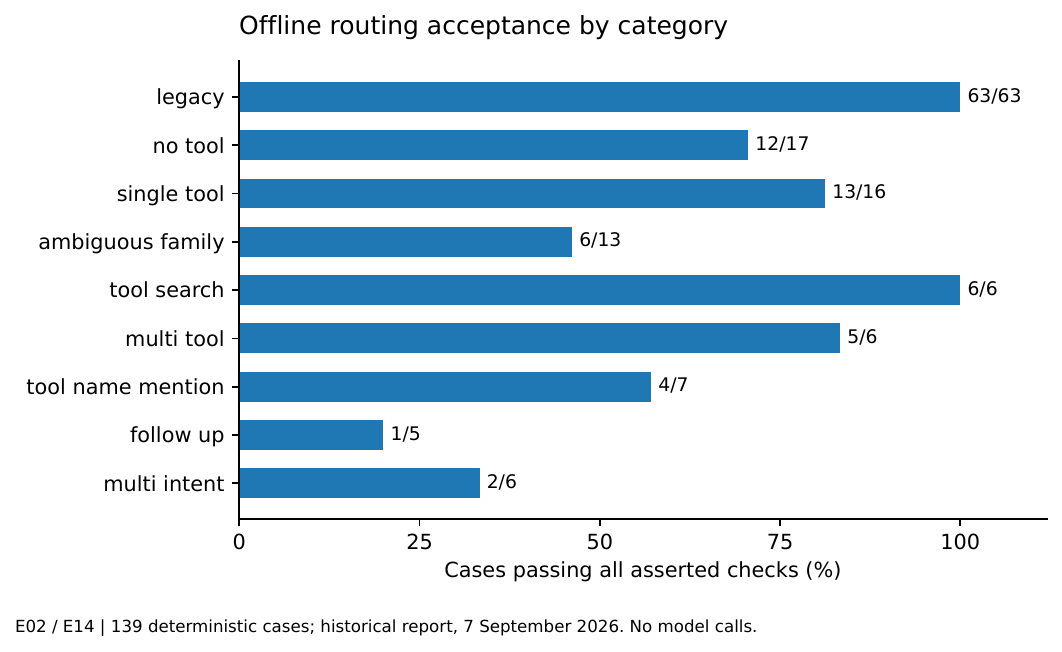}
\caption{Category-level acceptance counts and rates from the historical
offline report. The denominator is shown beside every bar. No model or
network execution occurs in this evaluation; no sampling-based error
bars are asserted.}
\end{figure}

Legacy cases contribute 63/139, or 45.32\%, of the corpus and pass
63/63. The remaining cases pass 49/76, or 64.47\%. The 16.10
percentage-point difference between the pooled and nonlegacy rates is a
composition identity, not a measured improvement. The smallest category
results, such as follow-up 1/5, should not be ranked as general
capability estimates.

The other overall values reconstruct to 33/40 for the implementation's
\nolinkurl{noToolPrecision}, 135/155 for required-tool recall, 16/216
for forbidden-tool admission, and 6/7 for fast-path accuracy. The first
label does not denote conventional prediction precision: its denominator
is explicitly no-tool cases with forbidden assertions. Unforbidden
discovery/UI tools are not scored as violations. The mean exposed-tool
count is reported as 11.73, and the mean schema-token estimate as
561.64. The latter omits unresolved definitions; its stated ``floor''
concerns the partial schema estimator, not an independently validated
lower bound on billed provider tokens (E02--E03).

The report lists three routing and 36 scoping failure events, with zero
budget events. These 39 events are not 39 failed cases; cases may have
multiple failed assertions. We retain both denominators rather than
forcing them to match.

\subsection{Isolated scoring
conformance}\label{isolated-scoring-conformance}

All 13 tests on the original metric function pass. They establish the
limited properties in Table 3, not the correctness of the whole pipeline
or any rate of real-world task completion.

\begin{table}[!htbp]
\centering
\footnotesize
\caption{Newly executed software conformance tests (E13). These are
intentionally constructed test inputs, not collected agent trajectories.
Model calls: zero; stateless-registry calls: zero.}
\vspace{4pt}
\begin{tabular}{@{}>{\raggedright\arraybackslash}p{(\columnwidth - 4\tabcolsep) * \real{0.09}}
>{\raggedright\arraybackslash}p{(\columnwidth - 4\tabcolsep) * \real{0.81}}
>{\raggedright\arraybackslash}p{(\columnwidth - 4\tabcolsep) * \real{0.1}}@{}}

\toprule\noalign{}
\begin{minipage}[b]{\linewidth}\raggedright
Test
\end{minipage} & \begin{minipage}[b]{\linewidth}\raggedright
Contract exercised
\end{minipage} & \begin{minipage}[b]{\linewidth}\raggedright
Outcome
\end{minipage} \\
\midrule\noalign{}

SC01 & Empty rate denominators return null, not perfect scores & Pass \\
SC02 & No-tool fixture without forbidden assertions is excluded from its
rate & Pass \\
SC03 & Known-gap failure stays in case and tool-recall denominators &
Pass \\
SC04 & Always-on un-forbidden discovery tools do not create a scoping
violation & Pass \\
SC05 & Required and forbidden names have separate explicitly weighted
denominators & Pass \\
SC06 & Fast-path accuracy counts only explicitly pinned fixtures &
Pass \\
SC07 & Reported means use two-decimal rounding & Pass \\
SC08 & A passing case with a retained gap annotation still counts as a
known gap & Pass \\
SC09 & A forbidden admitted name with a scoping failure reduces the
no-tool clean rate & Pass \\
SC10 & A non-scoping failure does not become a forbidden-admission
failure & Pass \\
SC11 & Permuting cases preserves every aggregate & Pass \\
SC12 & Required-tool recall weights assertions rather than averaging
case recall & Pass \\
SC13 & Repeating identical fixture inputs doubles counts but leaves
ratios and means unchanged & Pass \\
\bottomrule
\end{tabular}
\end{table}

The separately implemented verifier agrees on 91 descriptive
timing-statistic cells across seven reported selections and on all eight
compaction comparisons. All 12 verifier tests pass. These counts measure
software checks, not independent experimental observations. Exact
payload-to-CSV identity also holds for every non-null frequency entry
(E15).

\subsection{Operational latency: lifecycle age in an execution-shaped
field}\label{operational-latency-lifecycle-age-in-an-execution-shaped-field}

The export represents \textbf{8,843 attempt rows}. Of these,
\textbf{8,395 have numeric duration values} and \textbf{448 are null}, a
missingness fraction of \textbf{5.07\%}. There are 409 recorded zeros
and no negative recorded durations. The separate timestamp query reports
no negative start/end intervals. The primary analysis retains all
non-null values, including zeros and saturation (E06--E07).

\begin{table}[!htbp]
\centering
\footnotesize
\caption{Recorded latency by authority/status (E06; reanalysis E14). All
duration columns are milliseconds. ``N'' counts non-null and nonnegative
durations, not independently verified execution timing. The pooled row
is not a summary of a common task or model workload.}
\vspace{4pt}
\begin{tabular}{@{}>{\raggedright\arraybackslash}p{(\columnwidth - 10\tabcolsep) * \real{0.19}}
>{\raggedleft\arraybackslash}p{(\columnwidth - 10\tabcolsep) * \real{0.08}}
>{\raggedleft\arraybackslash}p{(\columnwidth - 10\tabcolsep) * \real{0.06}}
>{\raggedleft\arraybackslash}p{(\columnwidth - 10\tabcolsep) * \real{0.21}}
>{\raggedleft\arraybackslash}p{(\columnwidth - 10\tabcolsep) * \real{0.21}}
>{\raggedleft\arraybackslash}p{(\columnwidth - 10\tabcolsep) * \real{0.25}}@{}}

\toprule\noalign{}
\begin{minipage}[b]{\linewidth}\raggedright
Stratum
\end{minipage} & \begin{minipage}[b]{\linewidth}\raggedleft
N
\end{minipage} & \begin{minipage}[b]{\linewidth}\raggedleft
Null
\end{minipage} & \begin{minipage}[b]{\linewidth}\raggedleft
p50 (ms)
\end{minipage} & \begin{minipage}[b]{\linewidth}\raggedleft
p95 (ms)
\end{minipage} & \begin{minipage}[b]{\linewidth}\raggedleft
p99 (ms)
\end{minipage} \\
\midrule\noalign{}

Pooled & 8,395 & 448 & 1,765.00 & 36,250.10 & 2,147,483,647.00 \\
Server / completed & 7,068 & 437 & 1,645.50 & 22,751.95 & 38,118.31 \\
Server / failed & 697 & 11 & 986.00 & 21,207.60 & 32,108.16 \\
Server / reconciliation & 35 & 0 & 657,672.00 & 881,293.00 &
890,877.36 \\
Client / completed & 240 & 0 & 3,195.00 & 22,612.60 & 43,708.48 \\
Client / failed & 355 & 0 & 456,744,052.00 & 2,147,483,647.00 &
2,147,483,647.00 \\
\bottomrule
\end{tabular}
\end{table}

\begin{figure}[!htbp]
\centering
\includegraphics[width=1\textwidth,height=\textheight]{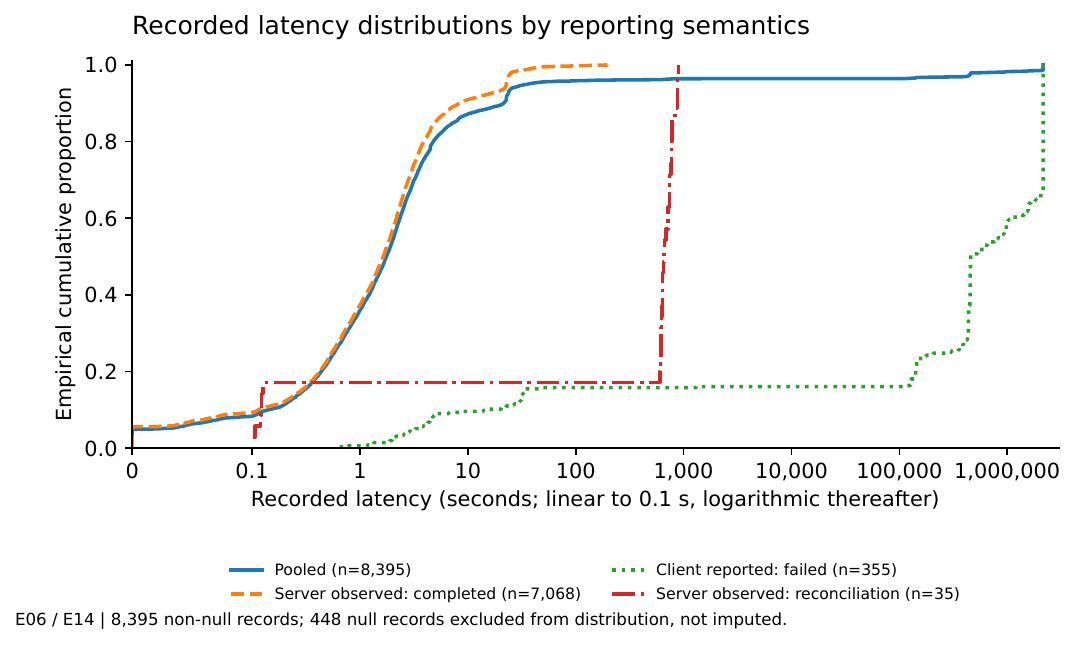}
\caption{Empirical cumulative distributions of recorded durations. The
selected strata illustrate different reporting semantics; all five
strata are included in Table 4 and the supplementary CSV. The horizontal
scale is linear near zero and logarithmic thereafter, preserving
zero-valued records and the large saturated tail.}
\end{figure}

The pooled mean is 43,538,604.60 ms, while the median is 1,765 ms. Its
99th percentile equals \textbf{2,147,483,647 ms}, the signed 32-bit
integer maximum. Exactly 121 values reach this cap, or 1.44\% of the
non-null records. Aggregate diagnostics place all 121 in the
client-reported, failed, \nolinkurl{client_never_reported} group. They
share a row-creation timestamp of 6 August 2026 at 15:15:25.168218 UTC
(E07). Row creation is not necessarily the time at which a tool began
execution.

The inspected current finalizer computes elapsed time between
finalization and the call's recorded start, floors it to milliseconds,
and clamps it to the integer cap. A retained migration defines a sweep
that finalizes abandoned
\nolinkurl{client_pending}/\nolinkurl{client_reported} calls as failed
with the \nolinkurl{client_never_reported} code (E08). The uncapped
timestamp intervals of the 121 saturated records range from
2,149,762,173.161 to 4,097,646,570.161 ms. This combination supports a
lifecycle-age interpretation rather than a claim that a provider
actively executed for approximately 24.86 days.

There are 298 \nolinkurl{client_never_reported} labels among 355
client-reported failures, including all saturated values. Merely
deleting the 121 capped records does not isolate execution timing: the
remaining pooled p99 is \textbf{453,532,616.08 ms}. Table 5 shows why
saturation removal is an insufficient correction.

\begin{table}[!htbp]
\centering
\footnotesize
\caption{Sensitivity to deleting only saturated values (E06; E14). This
is not an architecture comparison. Missing values are still excluded
from numeric summaries; all other lifecycle ages remain.}
\vspace{4pt}
\begin{tabular}{@{}>{\raggedright\arraybackslash}p{(\columnwidth - 8\tabcolsep) * \real{0.24}}
>{\raggedleft\arraybackslash}p{(\columnwidth - 8\tabcolsep) * \real{0.1}}
>{\raggedleft\arraybackslash}p{(\columnwidth - 8\tabcolsep) * \real{0.24}}
>{\raggedleft\arraybackslash}p{(\columnwidth - 8\tabcolsep) * \real{0.14}}
>{\raggedleft\arraybackslash}p{(\columnwidth - 8\tabcolsep) * \real{0.28}}@{}}

\toprule\noalign{}
\begin{minipage}[b]{\linewidth}\raggedright
Selection
\end{minipage} & \begin{minipage}[b]{\linewidth}\raggedleft
Numeric N
\end{minipage} & \begin{minipage}[b]{\linewidth}\raggedleft
Mean (ms)
\end{minipage} & \begin{minipage}[b]{\linewidth}\raggedleft
p50 (ms)
\end{minipage} & \begin{minipage}[b]{\linewidth}\raggedleft
p99 (ms)
\end{minipage} \\
\midrule\noalign{}

All recorded values & 8,395 & 43,538,604.60 & 1,765.00 &
2,147,483,647.00 \\
Cap excluded only & 8,274 & 12,770,251.91 & 1,722.00 & 453,532,616.08 \\
\bottomrule
\end{tabular}
\end{table}

Within server-observed completed records, the median is 1,645.5 ms and
p99 is 38,118.31 ms. This is a more narrowly defined descriptive
stratum, not an estimate of overall user-facing latency. Its 437 missing
values, 389 zeros, workload mix, potential test/backfill records, and
unobserved model versions remain relevant. The contrast with the pooled
p99 is not a speedup attributable to an engineering change.

Missingness remains material inside that stratum: 437 of 7,505
server-observed completed rows lack a duration. At the illustrative
30-second threshold, 6,967 of 7,068 observed durations are at or below
the threshold (98.57\%). Over every completion of the 437 missing
threshold classifications, the all-row fraction ranges from 6,967/7,505
to 7,404/7,505 (92.83\%--98.65\%; E15). No value is imputed and neither
endpoint is a population confidence limit. This calculation concerns the
recorded field, not verified active execution or a service-level
objective.

Finally, 130 rows have a difference exceeding one second between
recorded latency and the queried start/end interval; 121 are the
saturated rows. The other nine differences are unresolved by the
retained aggregates. We neither silently correct them nor assign a
cause. All 8,843 inspected owner/call-key combinations are distinct, but
this does not rule out retries under different keys, duplicate logical
tasks, or within-user dependence.

\subsection{Compaction: follow-up savings and two-call
accounting}\label{compaction-follow-up-savings-and-two-call-accounting}

The historical compaction document describes one paired trajectory with
25 history turns. It reports a \nolinkurl{gpt-5.6} model identifier,
\nolinkurl{ai} 7.0.59 and its \nolinkurl{@ai-sdk/openai} adapter 4.0.37;
an immutable provider-model snapshot, temperature, seed, and hardware
description were not established from the retrieved evidence (E04). The
document's reported observations are shown in Table 6.

\begin{table}[!htbp]
\centering
\footnotesize
\caption{Documented compaction-pilot values and derived sums (E04; E14).
There is one paired trajectory, not a population sample. Positive
relative changes mean a larger native value. The two duration sums are
arithmetic, not independently measured end-to-end wall
times.}
\vspace{4pt}
\begin{tabular}{@{}lrrr@{}}

\toprule\noalign{}
Quantity & Application & Native & Change (\%) \\
\midrule\noalign{}

Trigger input (tokens) & 10,610 & 10,790 & +1.70 \\
Follow-up input (tokens) & 10,695 & 600 & -94.39 \\
Two-call input sum (tokens) & 21,305 & 11,390 & -46.54 \\
Trigger TTFT (ms) & 1,179 & 2,016 & +70.99 \\
Follow-up TTFT (ms) & 922 & 974 & +5.64 \\
Trigger full turn (ms) & 1,648 & 2,474 & +50.12 \\
Follow-up full turn (ms) & 1,254 & 1,230 & -1.91 \\
Two-call duration sum (ms) & 2,902 & 3,704 & +27.64 \\
\bottomrule
\end{tabular}
\end{table}

\begin{figure}[!htbp]
\centering
\includegraphics[width=0.9\textwidth,height=\textheight]{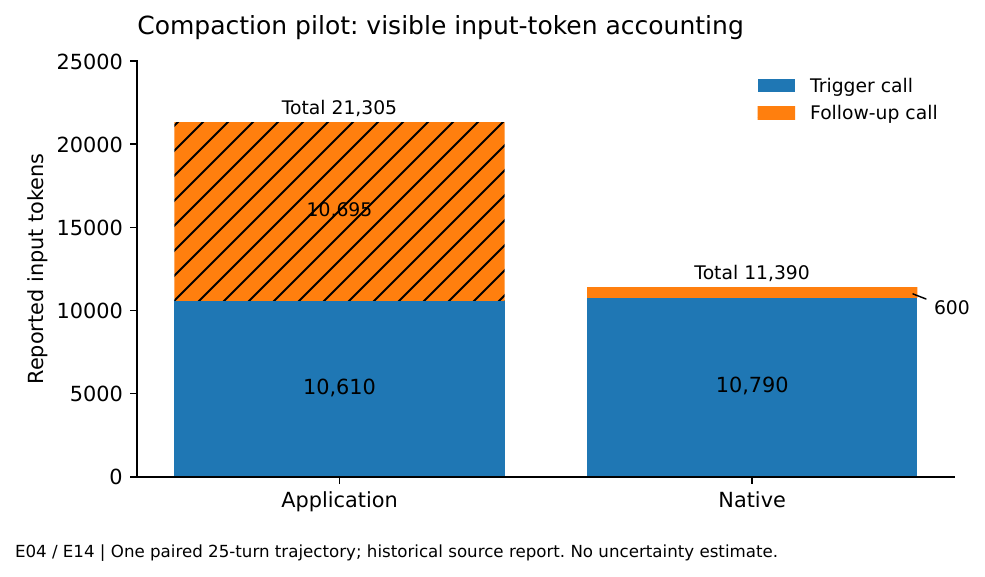}
\caption{Input-token accounting for the two calls exposed in the
historical compaction report. Including the trigger changes the
denominator of the comparison; it does not estimate complete provider
billing or the full 25-turn trajectory cost.}
\end{figure}

The follow-up input count falls from 10,695 to 600, a 94.39\% reduction.
Including the trigger yields 21,305 application-side input tokens and
11,390 native-side input tokens, a 46.54\% reduction. Neither result is
false as arithmetic. The smaller value answers a different, broader
question, but it still excludes any unreported compaction-only usage and
the rest of the trajectory. Total cost savings are \textbf{not
established}.

The native trigger has 10,790 input tokens, 2,598 above the reported
configured 8,192-token limit. The source attributes this to
underestimation by its bytes/4 estimator and explicitly states that the
run does not qualify request-envelope enforcement (E04). That
source-reported qualification failure must remain visible. It is not a
claim that a provider's hard model-context ceiling was exceeded; the
configured application limit and the provider ceiling are different
quantities.

\begin{figure}[!htbp]
\centering
\includegraphics[width=0.9\textwidth,height=\textheight]{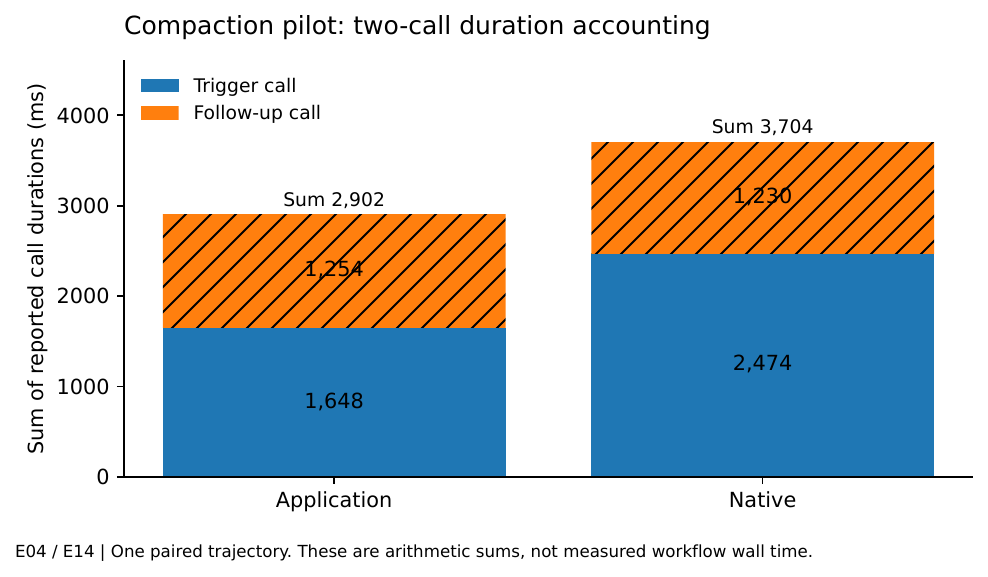}
\caption{Sum of the trigger and follow-up full-turn durations reported
in the pilot. The native sum is larger even though its follow-up
duration is slightly smaller. No repeated-run uncertainty or total
workflow duration is available.}
\end{figure}

The follow-up full-turn duration changes from 1,254 to 1,230 ms, while
the trigger changes from 1,648 to 2,474 ms. Their sums are 2,902 and
3,704 ms, respectively: a 27.64\% increase. Summation does not establish
an actual workflow wall-clock duration, because orchestration boundaries
and additional work are not supplied. It does preclude treating the
follow-up-only timing as complete two-call accounting.

Both arms reportedly pass three ordinary-fact and two safety-state
checks. Budget information and the two safety states were re-injected by
the application; only two native-context facts were independently probed
without that reinjection. The checks therefore do not isolate retention
of all asserted facts by the native compaction mechanism. The document
also reports persistence/reload behavior but does not supply a measured
database or object-storage overhead. These limitations are retained
rather than folded into a general memory-quality claim.

\subsection{Secondary historical evidence and absent
comparisons}\label{secondary-historical-evidence-and-absent-comparisons}

The discovery pilot uses one weather task with two executions per arm.
Initial schema sizes are reported as 1,112 and 1,108 bytes, with
byte-divided-by-four estimates of 278 and 277 tokens. Reported
answer-TTFT p50 values are 3,008 and 3,264 ms; p95 values are 3,115 and
4,082 ms. Required-read and required-fact checks pass 2/2 in both arms.
With this limited documentary sample and no raw timings, superiority,
equivalence, and tail-latency claims are not established. The complete
reported discovery summary is retained in CSV, including full-turn
percentiles and zero observed cache-read counts as a configuration note
(E04).

A seven-tool deterministic ledger-outage artifact reports mutation
execution despite a failed durable claim in 7/7 baseline fixtures and
0/7 candidate fixtures. The candidate is explicitly named uncommitted,
and the artifact marks product-level comparison as not tested (E05).
Recomputed Wilson endpoints agree with the reported values, but the
interval calculation is not evidence that these curated fixtures form
independent draws from future failure conditions. We do not attach a
paired significance claim: fixture selection, pairing provenance, and a
sampling or randomization design were not verified. The endpoint counts
alone cannot justify extrapolation to production safety.

The selected database inventory contains zero rows in
\nolinkurl{platform_experiments}, four pass-labeled deployed evaluation
rows, and zero rows in each of two other selected prediction/evaluation
tables. The episode score-source counts are 1,173
\nolinkurl{turn_judge}, 61 \nolinkurl{verifier}, and 77
\nolinkurl{critique} (E09). These are record labels, not independently
adjudicated success counts. The inventory does not justify declaring
that no A/B tests exist elsewhere, nor does it supply a controlled
current-pipeline comparison.

\section{Statistical interpretation and attempted
falsification}\label{statistical-interpretation-and-attempted-falsification}

The main results are exact descriptive statements about finite retained
artifacts. Their principal uncertainty is evidential and semantic, not a
confidence interval around a random sample mean. The curated routing
corpus has no demonstrated random-sampling mechanism; tool-name
assertions within one case share context. Operational records may share
users, tasks, providers, deployments, or retries. The compaction pilot
has one trajectory and mixes memory with explicit reinjection.
Resampling these objects as independent trials would add an assumption
the evidence does not justify.

Accordingly, the main results contain no p-values, standardized
treatment effect sizes, or population confidence intervals. Relative
changes and percentage-point differences are fully reported as
arithmetic contrasts. A claim of statistical significance is not
established. Conversely, an absent significance result is not evidence
of equivalence or zero effect. A numerical power requirement for a
future paired trial would need a prespecified meaningful effect,
discordant-outcome rates or trajectory-level variance, dependence
structure, and error-rate targets. None is supplied as an observed
property of this dataset.

We actively checked alternative explanations. First, the apparently
perfect development gate is explained by known-gap policy; it does not
require a miscounted case denominator. Second, saturated timing values
are corroborated by current clamp logic and abandoned-call labels,
rather than discarded as unspecified outliers. Third, removing the cap
alone leaves extreme lifecycle ages, weakening the hypothesis that a
handful of corrupt values is the sole problem. Fourth, token savings are
recomputed at both visible reporting boundaries. Fifth, reinjected facts
are not counted as isolated evidence of compaction retention. Sixth, the
separate weighted-arithmetic implementation checks the expanded-array
calculations. Seventh, all selected missingness thresholds are reported,
including bounds that weaken a complete-case reading. Passing verifier
rejection tests supports error detection in the analysis code; it is not
a substitute for independent source attestation or external peer review.

Several falsification attempts remain incomplete. We could not establish
that every operational row came from production-only traffic,
reconstruct historical procedure deployments, identify task-level
duplication, measure provider cache effects beyond the pilot's reported
counts, or validate the historical model identifier against an immutable
provider snapshot. Those gaps constrain the conclusions. The audit is
evidence against unsupported interpretations, not proof that every
unfavorable observation reflects a defect in the agent itself.

\section{Discussion: a proposed measurement
contract}\label{discussion-a-proposed-measurement-contract}

The three cases share a structure: a local operational metric is
promoted to a broader scientific construct without preserving its
boundary. Gate status concerns a release policy, not necessarily case
correctness. An elapsed-age column concerns the interval between
recorded events, not necessarily active execution. A follow-up input
count concerns one provider request, not a billing-complete workflow.
The data illustrate these distinctions; they do not establish how often
the same problem occurs in other systems.

A useful corrective is to make the measurement unit and event semantics
explicit before aggregation. We propose a contract containing the
analysis unit, eligibility rule, model/prompt/scorer/deployment
revisions, start and end events, clock source, reporting authority,
terminal status, censoring state, independent outcome evidence, per-call
usage, and task/trajectory grouping. The supplemental JSON gives a
field-level proposal. It is not a claim that production currently
implements this contract or that its adoption has been experimentally
shown to improve reliability.

For time measurement, execution, queueing, approval waiting,
reconciliation, and abandonment age should be distinct event types. Wide
numeric storage and explicit saturation flags preserve the difference
between a large observed interval and a clipped value. A completed tool
call should not imply a verified task outcome. A reconciliation event
should not be silently relabeled as either task failure or active
execution time.

For acceptance reporting, case outcomes, mandatory-gate outcomes, and
assertion-level rates should be published together. Exemptions can be
legitimate engineering policy, but they should remain visible in
scientific denominators. Previously developed fixtures and newly added
categories should be identified; their mixture should not be mistaken
for a treatment effect.

For compaction, the primary accounting boundary should be stated before
inspection of results. A billing-oriented endpoint needs all relevant
provider receipts, including input, output, cache, retries, and
separately accounted compaction where exposed. A latency-oriented
endpoint needs a measured start-to-terminal boundary rather than a
selective call duration. Retention checks should identify which facts
remain in native context and which are deliberately reintroduced.
Reinjection may be desirable system design; it simply answers a
different mechanism question.

These recommendations overlap existing evaluation and evidence-review
ideas \citep{kapoor2024, ren2026}. The present contribution is the
reproducible, source-linked demonstration of their relevance to one
system. The proposed contract's effectiveness remains a hypothesis to
test prospectively. The revision adds an executable verifier for the
supplied frequency artifacts, not a production implementation of that
contract. Neither the application nor its database schema was changed.

\section{Controlled follow-up studies: not
run}\label{controlled-follow-up-studies-not-run}

The ancillary \nolinkurl{operator_coverage} directory supplies 264 proposed
scenario contracts in 27 families, 40 fault overlays, and a 36-case smoke
selection. All scenarios are marked \nolinkurl{NOT_RUN}. This prospective
catalog contributes no executed task outcomes or performance estimates
to the present audit.

The following studies are necessary for stronger claims. They are
specifications for future execution, not completed experiments,
preregistered studies, or fabricated sample-size commitments. A disjoint
calibration set and prespecified effect criterion must determine sample
size before the evaluation set is opened. The task, trajectory, or
fault-matrix cell must be the declared analysis unit; individual fact
assertions and repeated software checks cannot inflate it. Invalid
infrastructure runs require prespecified rules; unsuccessful agent
outcomes cannot simply be removed.

\textbf{P0: Validate timing semantics in a sandbox.} Exercise immediate
completion, provider delay, approval waiting, abandoned-client
finalization, reconciliation, and values around the integer-storage
boundary under a controlled clock. Compare each reported field with an
independently retained event interval. The null hypothesis is that the
proposed revised instrumentation does not reduce semantic
misclassification relative to the frozen current implementation; the
alternative is a reduction on the locked event matrix. First establish
deterministic event conformance, then evaluate any operational
misclassification rate with blinded labels. Stop at the fixed matrix and
planned repetitions, not after a favorable subset.

\textbf{P1: Ablate current-path tool admission.} Compare the current
dynamic admission mechanism with a static-tool baseline using the same
frozen executing model, provider, prompts, tasks, tool implementations,
retry policy, timeout, and resource budget. Only the admission mechanism
should differ. Preserve task pairing and counterbalance execution order.
The primary outcome should be independently verified task completion;
admitted schema size, provider tokens, false-success reports, safety
violations, and measured user-facing duration are secondary. For a
prespecified target task distribution, the null is zero paired
completion difference; a directional alternative is appropriate only if
committed before evaluation. Task-cluster resampling or a paired
categorical test must reflect the actual repeated-trial design. Correct
a declared secondary-test family, rather than selecting whichever metric
improves.

\textbf{P2: Evaluate full-trajectory compaction.} Use paired histories
with a locked horizon distribution and identical model, prompt,
reinjection, tool, and budget policies. Separate native-context recall,
application-restored state, safety decisions, and verified downstream
outcomes. The primary estimand should be a prespecified efficiency
measure conditional on a prespecified quality requirement, or a joint
quality/cost comparison; the margin must be chosen before evaluation.
The null for a cost-reduction claim is no reduction in billing-complete
trajectory cost under the quality requirement. Record all provider
receipts and wall-clock boundaries. Analyze trajectories as clusters,
not individual fact checks or turns. A quality-preserving efficiency
conclusion requires evidence for both parts, not a significant token
difference alone.

\textbf{P3: Validate durable effects under faults.} Freeze both baseline
and candidate revisions. Execute an independent fault matrix covering
durable-claim failure, duplicate requests, lost responses after commit,
revoked permission, and interruption at defined boundaries. Use a
provider simulator or explicitly authorized sandbox, never unapproved
production mutations. Read the external state independently and count
missing, duplicate, forbidden, and authorized effects. A claim that
fail-closed admission reduces forbidden effects has a null of no
reduction on the specified fault distribution. Deterministic matrix
coverage and stochastic repetitions should be reported separately, with
all outcomes and retries preserved.

\textbf{P4: Obtain independently verified task outcomes.} Lock a
held-out corpus of realistic tasks and publish inclusion criteria,
model/tool availability, verifier rules, and scorer revision. Prevent
use of held-out failures for candidate selection. Compare against a
matched conventional loop or mechanism-disabled baseline rather than an
unrelated leaderboard entry. Calibrate any automated judge against
blinded human or state-based adjudication and retain disagreements. The
primary hypothesis concerns verified completion, not the count of
records whose status field says completed. Select the final candidate
before opening the holdout; report every planned arm and run.

No numerical power claim or minimum sample-size guarantee is made for
these studies. Their required observations, controls, stopping rules,
and statistical choices depend on target populations and calibration
evidence not present in the current package.

\section{Limitations and threats to
validity}\label{limitations-and-threats-to-validity}

\textbf{Internal validity.} The audit links database code inspected
during extraction with historical operational records, but it does not
reconstruct every historical deployment. Matching labels, timestamps,
and clamp behavior support a lifecycle interpretation without proving
the complete causal history. The provider pilot changes compaction
behavior and includes reinjection; its small reported comparison does
not isolate every mechanism. The artifact selection was retrospective
and focused on accessible evidence, not sampled independently of the
emerging thesis.

\textbf{Construct validity.} Routing admission is not task execution;
completed attempt status is not verified user-goal completion; a named
verifier source is not independent adjudication. The schema-token floor
is a partial estimator. The latency field measures a stored interval
with mixed event semantics. We narrow claims to these constructs rather
than substitute better-sounding labels.

\textbf{Statistical conclusion validity.} Independence, sampling
coverage, task clusters, and full repeated-run distributions are
unavailable. The single compaction trajectory cannot support variance
estimation for a target population. The historical seven-fixture
interval calculation does not resolve dependence or selection. We
therefore avoid inferential conclusions rather than presenting
misleading certainty. The additional missingness bounds constrain only
possible completions of the finite table; they do not repair workload
selection, distinguish active execution from waiting, or restore missing
task clusters.

\textbf{External validity.} One private agent system and selected
internal artifacts do not establish prevalence across frameworks,
providers, models, or task populations. The offline routing report
predates the currently documented path. No standardized task-performance
score or fair external system comparison was independently reproduced.
We make no generalization to enterprise reliability, world-model
learning, adaptive prediction, multi-agent superiority, or autonomous
scientific discovery.

\textbf{Reproducibility validity.} The frequency multiset and derived
figures are reproducible offline, but the source database is mutable and
its extracts are non-atomic. The two numerical implementations share
those same inputs, so agreement can catch implementation discrepancies
but not a common source error. Full historical provider receipts and
acceptance case records are not packaged. Original repository access
remains private, and the isolated TypeScript conformance test is not an
environment-identical application rerun. A source hash is useful
evidence of file identity; it is not proof of public accessibility or
execution provenance.

\section{Reproducibility, ethics, and
availability}\label{reproducibility-ethics-and-availability}

The companion archive contains the evidence registry, claims ledger,
experiment registry, traceability map, exact latency frequencies and
compact payload digests, historical source-derived aggregates, the
unmodified metric-function source, conformance tests, analysis/figure
scripts, SQL query templates, environment receipts, a second arithmetic
verifier, and missingness bounds. The principal commands are:

\noindent\begin{minipage}{\linewidth}
\begin{verbatim}
python analysis/reanalyse.py
node analysis/scoring_conformance.cjs
python analysis/review_validation.py
\end{verbatim}
\end{minipage}

The Node command requires the recorded TypeScript dependency; the README
explains resolution. The historical provider and complete routing
commands are documented separately as \textbf{not executed here}.
Re-running the read-only SQL against a later database state is a new
extract, not reproduction of the retained snapshot window. Offline
reproduction should begin with the packaged data and checksum
validation. Checksums protect file identity, not factual correctness or
anonymity. The submission-oriented source package includes
research-supporting data and code, the editable manuscript source,
execution receipts, and the separately labelled prospective operator
coverage catalog in its ancillary directory.

No protected prompts, credentials, raw account identifiers, or
conversation content are included. Aggregation reduces direct identifier
exposure but does not prove anonymity. The ancillary files retain exact
operational frequency information and selected source excerpts; they do
not disclose the complete private application repository.

This manuscript and its local analysis code were prepared and reviewed
with AI assistance. The additional passes are computational and
editorial verification within the same preparation workflow, not reviews
by independent human scientists. The human authors are responsible for
the manuscript and its claims. The proposed experiments must use isolated environments for
potentially destructive effects and preserve approval/tenant boundaries.
We do not offer the retrospective observations as an authorization to
weaken those controls.

\section{Conclusion}\label{conclusion}

The available Praxa evidence supports a measurement case study, not a
general superiority claim. A historical routing gate has zero gating
failures while 27 of 139 cases fail. An operational duration field
contains capped ages of abandoned client calls, making its pooled tail
unsuitable as a claim about active execution latency. A documented
compaction pilot's token reduction changes from 94.39\% at the follow-up
boundary to 46.54\% across the two visible calls, without establishing
total cost savings.

These findings are accompanied by reproducible aggregate calculations,
13 passing isolated scoring-function tests, and 12 passing verifier
tests. A separate weighted-arithmetic implementation confirms 91
timing-statistic cells and all eight reported compaction comparisons;
finite missingness bounds preserve what remains unresolved. Their value
lies in identifying exactly what each result measures and what
additional evidence would be needed to interpret it more broadly. A
stronger current-pipeline capability paper remains contingent on frozen,
controlled, independently verified task experiments. The present package
makes that boundary explicit and provides the data and procedures needed
to challenge its narrower conclusions.

\section*{Appendix A. Evidence identity and source
coverage}\label{appendix-a.-evidence-identity-and-source-coverage}
\addcontentsline{toc}{section}{Appendix A. Evidence identity and source
coverage}

The CSV registry is the authoritative full mapping. Repository source
paths are interpreted at E01 unless a separate historical revision is
stated. The following identities support the central implementation and
historical artifacts:

\textbf{E02.}
\nolinkurl{evals/acceptance/agentic-baseline-2026-09-07.json}. Git blob:
\nolinkurl{9925e82a6e45855af2670b43adb20c9a10c97d3e}.

\textbf{E03.} \nolinkurl{src/lib/eval/agentic-metrics.ts}. Git blob:
\nolinkurl{01bf2308619a9c9b4a84f7ce5b6803d3942b6498}.

\textbf{E04.}
\nolinkurl{docs/benchmarks/native-quality-pilot-2026-09-06.md}. Git
blob: \nolinkurl{075810098d19906b75d75bc3723591f18c3817a9}.

\textbf{E05.} \nolinkurl{artifacts/frontier/comparison-results.json}.
Git blob: \nolinkurl{fa12523278918db48764b9b21b573f99e9a02aee}.

\textbf{E10.} \nolinkurl{evals/scoring/frontier-outcome-scorer.ts}. Git
blob: \nolinkurl{15e390c206d094a775b71ba2dd95e60014113f78}.

\textbf{E10.} \nolinkurl{evals/frontier/contracts.ts}. Git blob:
\nolinkurl{4eb2bef13c0c8c3795e0642e994e92cd7f3cda13}.

Database metadata has an independent identity. The inspected
\nolinkurl{finalize_agent_tool_call} source is reported by PostgreSQL as
4,979 characters with MD5 \nolinkurl{5ce45a9ba6e845bde047f7e1d9ac6120};
the client finalizer is 763 characters with MD5
\nolinkurl{ce7c0b30c2e14f004b65b2cc04b0dbd3}. These are returned
identities of complete source strings, not hashes of the selected
excerpt file. The relevant inspected expression is:

\noindent\begin{minipage}{\linewidth}
\begin{verbatim}
measured_latency := least(
  2147483647::numeric,
  greatest(
    0::numeric,
    floor(extract(epoch from
      (finished_time - current_call.started_at)) * 1000)
  )
)::integer;
\end{verbatim}
\end{minipage}

The associated migration metadata identifies version
\nolinkurl{20260806140000}, named
\nolinkurl{sweep_abandoned_client_tool_calls}. Its source specifies a
default stale horizon of 86,400 seconds and default limit of 200. We
read that definition but did not call it. Historical deployment linkage
remains unresolved.

\section*{Appendix B. Transparent calculations and interval
status}\label{appendix-b.-transparent-calculations-and-interval-status}
\addcontentsline{toc}{section}{Appendix B. Transparent calculations and
interval status}

The arithmetic underlying the abstract and central tables is:

\[
\begin{aligned}
\text{case pass rate}&=112/139=0.8057554,\\
\text{nonlegacy pass rate}&=(112-63)/(139-63)=49/76,\\
\text{missing fraction}&=448/8843=0.0506615,\\
\text{follow-up reduction}&=1-600/10695=0.9438990,\\
\text{two-call input reduction}&=1-(10790+600)/(10610+10695)\\
&=0.4653837,\\
\text{duration-sum change}&=(2474+1230)/(1648+1254)-1\\
&=0.2763611.
\end{aligned}
\]

The historical ledger artifact's Wilson interval can be reproduced
algebraically. For \(\hat p=k/n\) and \(z=1.959963984540054\), its
center and half-width are

\[
c=\frac{\hat p+z^2/(2n)}{1+z^2/n},\qquad
w=\frac{z\sqrt{\hat p(1-\hat p)/n+z^2/(4n^2)}}{1+z^2/n}.
\]

For \(k=n=7\), the endpoints are approximately {[}0.64567, 1{]}; for
\(k=0,n=7\), they are {[}0, 0.35433{]}. This reproduces the source's
interval calculation, not its inferential applicability. A conventional
binomial interpretation would require conditions that were not
established for the selected deterministic fixtures. The main paper
therefore does not use these intervals as confidence statements about
production safety.

\section*{Appendix C. Experimental settings and missing
fields}\label{appendix-c.-experimental-settings-and-missing-fields}
\addcontentsline{toc}{section}{Appendix C. Experimental settings and
missing fields}

\begin{table}[!htbp]
\centering
\footnotesize
\caption{Reproducibility settings and unresolved fields. An omitted
field is not filled with a conventional default. A model name reported
by an artifact is not independently verified provider
metadata.}
\vspace{4pt}
\begin{tabular}{@{}>{\raggedright\arraybackslash}p{(\columnwidth - 4\tabcolsep) * \real{0.2}}
>{\raggedright\arraybackslash}p{(\columnwidth - 4\tabcolsep) * \real{0.4}}
>{\raggedright\arraybackslash}p{(\columnwidth - 4\tabcolsep) * \real{0.4}}@{}}

\toprule\noalign{}
\begin{minipage}[b]{\linewidth}\raggedright
Analysis
\end{minipage} & \begin{minipage}[b]{\linewidth}\raggedright
Recorded setting
\end{minipage} & \begin{minipage}[b]{\linewidth}\raggedright
Missing or not established
\end{minipage} \\
\midrule\noalign{}

Historical routing & 139 curated cases; source generation time and base
SHA; no model/network & Complete raw-case package, clean historical
tree, environment-identical replay \\
Scoring conformance & Original source hash; Node 22.16.0; TypeScript
5.8.3; 13 tests & Full project typecheck and end-to-end route
execution \\
Operational timing & Exact authority/status frequencies; extraction
window; null counts & Atomic snapshot, task clusters, model versions,
workload eligibility and complete retention \\
Native pilot & Reported model \nolinkurl{gpt-5.6}; \nolinkurl{ai}
7.0.59; \nolinkurl{@ai-sdk/openai} 4.0.37; documented task counts &
Immutable model snapshot, raw receipts, seed, temperature, hardware and
complete billing \\
Ledger outage & Seven fixtures/arm; named historical baseline; candidate
marked uncommitted & Candidate SHA, paired raw cases, product-path
replay \\
\bottomrule
\end{tabular}
\end{table}

The data dictionary identifies each CSV's unit and whether it contains
retrieved values or derived calculations. In particular,
\nolinkurl{latency_frequencies.csv} is a compressed multiset, not one
row per call; \nolinkurl{acceptance_categories.csv} contains aggregate
categories, not independent trials; and the pilot CSVs transcribe
historical reports rather than encode new experiments. The validation
receipts distinguish the newly executed tests and descriptive checks
from all unrun follow-up work.

\section*{Appendix D. Finite missingness sensitivity and
verification}\label{appendix-d.-finite-missingness-sensitivity-and-verification}
\addcontentsline{toc}{section}{Appendix D. Finite missingness
sensitivity and verification}

The table below reports the complete illustrative threshold set.
``Observed-only'' divides the observed count at/below the threshold by
the non-null count. The completion range instead retains all rows in its
denominator and varies every missing threshold classification. The
pooled selection has 8,395 observed and 448 missing durations; the
server/completed selection has 7,068 observed and 437 missing. Values
are percentages. No missing value is set to zero in the primary
analysis.

\begin{table}[!htbp]
\centering
\footnotesize
\caption{Finite-completion bounds for recorded-duration thresholds (E06;
E15). These are not confidence intervals, production SLO estimates, or
measurements of active execution. The thresholds were chosen for
descriptive sensitivity analysis, not before data collection. Exact
numerators and rational endpoints are in
\nolinkurl{missingness_sensitivity_bounds.csv}.}
\vspace{4pt}
\begin{tabular}{@{}>{\raggedright\arraybackslash}p{(\columnwidth - 8\tabcolsep) * \real{0.22}}
>{\raggedleft\arraybackslash}p{(\columnwidth - 8\tabcolsep) * \real{0.12}}
>{\raggedleft\arraybackslash}p{(\columnwidth - 8\tabcolsep) * \real{0.2}}
>{\raggedleft\arraybackslash}p{(\columnwidth - 8\tabcolsep) * \real{0.2}}
>{\raggedleft\arraybackslash}p{(\columnwidth - 8\tabcolsep) * \real{0.26}}@{}}

\toprule\noalign{}
\begin{minipage}[b]{\linewidth}\raggedright
Selection
\end{minipage} & \begin{minipage}[b]{\linewidth}\raggedleft
Threshold (s)
\end{minipage} & \begin{minipage}[b]{\linewidth}\raggedleft
Observed count
\end{minipage} & \begin{minipage}[b]{\linewidth}\raggedleft
Observed-only (\%)
\end{minipage} & \begin{minipage}[b]{\linewidth}\raggedleft
Completion range (\%)
\end{minipage} \\
\midrule\noalign{}

Pooled & 1 & 3,016 & 35.93 & 34.11--39.17 \\
Pooled & 5 & 6,772 & 80.67 & 76.58--81.65 \\
Pooled & 30 & 7,936 & 94.53 & 89.74--94.81 \\
Pooled & 60 & 8,030 & 95.65 & 90.81--95.87 \\
Server/completed & 1 & 2,638 & 37.32 & 35.15--40.97 \\
Server/completed & 5 & 6,010 & 85.03 & 80.08--85.90 \\
Server/completed & 30 & 6,967 & 98.57 & 92.83--98.65 \\
Server/completed & 60 & 7,037 & 99.56 & 93.76--99.59 \\
\bottomrule
\end{tabular}
\end{table}

The verifier evaluates weighted quantiles directly from ordered distinct
values and frequencies. Its mean uses the integer sum
\(S_1=\sum_v vf(v)\); its variance uses \([nS_2-S_1^2]/[n(n-1)]\), where
\(S_2=\sum_v v^2f(v)\). Rational arithmetic avoids accumulation error
before conversion for comparison. The verifier does not import NumPy,
call a model, or access the source database. Twelve constructed tests
include a same-total/different-distribution corruption that must be
rejected: agreeing totals alone would not detect that error. The
supplied source data are never modified by these tests.

\nolinkurl{validation/second_pass_validation.json} records the exact
rational checks, test outcomes, and execution environment. The upload
review's rerun receipts are retained separately under
\nolinkurl{validation/upload_review_2026-09-10/}. Neither a new check count nor a
regenerated execution timestamp creates another empirical agent
observation.
\bibliographystyle{unsrtnat}
\bibliography{references}
\end{document}